\documentclass[twocolumn,trackchanges]{aastex631}
\usepackage{natbib}
\usepackage{epsfig}
\usepackage{graphicx}
\usepackage{subfigure}
\usepackage{float}
\usepackage{amsmath}
\usepackage{color}
\usepackage{amssymb}
\usepackage{apjfonts}

\newcommand{\PMO}{Purple Mountain Observatory, Chinese Academy of Sciences, Nanjing 210023, China}
\newcommand{\USTC}{School of Astronomy and Space Sciences, University of Science and Technology of China, Hefei 230026, China}

\begin{document}

\title{New Tests on Lorentz Invariance Violation Using Energy-Resolved Polarimetry of Gamma-Ray Bursts}

\correspondingauthor{Jun-Jie Wei}
\email{jjwei@pmo.ac.cn}

\author[0000-0003-0162-2488]{Jun-Jie Wei}
\affiliation{\PMO}
\affiliation{\USTC}

\begin{abstract}
One of the manifestations of Lorentz invariance violation (LIV) is vacuum birefringence,
which leads to an energy-dependent rotation of the polarization plane of linearly polarized photons
arising from an astrophysical source. Here we use the energy-resolved polarization measurements
in the prompt $\gamma$-ray emission of five bright gamma-ray bursts (GRBs) to constrain this vacuum birefringent
effect. Our results show that at the 95\% confidence level, the birefringent parameter $\eta$
characterizing the broken degree of Lorentz invariance can be constrained to be $|\eta|<\mathcal{O}(10^{-15}-10^{-16})$,
which represent an improvement of at least eight orders of magnitude over existing limits from
multi-band optical polarization observations. Moreover, our constraints are competitive with
previous best bounds from the single $\gamma$-ray polarimetry of other GRBs. We emphasize that,
thanks to the adoption of the energy-resolved polarimetric data set, our results on $\eta$ are
statistically more robust. Future polarization measurements of GRBs at higher energies and
larger distances would further improve LIV limits through the birefringent effect.
\end{abstract}


\section{Introduction}
\label{sec:intro}
Lorentz invariance is a fundamental postulate of Einstein's theory of relativity as well as of the Standard Model
of particle physics. However, a violation or deformation of Lorentz invariance at around the Planck energy scale
$E_\mathrm{Pl}=\sqrt{\hbar c^{5}/G}\simeq1.22\times10^{19}$ GeV are predicted in many quantum gravity (QG) theories
attempting to unify quantum theory and gravity
\citep{1989PhRvD..39..683K,1997IJMPA..12..607A,1999GReGr..31.1257E,2002Natur.418...34A,2002PhRvL..88s0403M,
2002PhRvD..65j3509A,2005LRR.....8....5M,2009PhLB..679..407L,2013LRR....16....5A,2014RPPh...77f2901T,
2021FrPhy..1644300W,2022Univ....8..323H,2022PrPNP.12503948A,Wei2022,2023arXiv231200409A,Desai2024}.
Experimental searches for potential signatures of Lorentz invariance violation (LIV) have been conducted
in various systems (see \citealt{2011RvMP...83...11K} for a continuously updated compilation).

For photons with energy $E\ll E_\mathrm{Pl}$, one of the most discussed LIV-induced modifications to the
dispersion relation can be approximated by a Taylor series, retaining only the first leading order term \citep{2003PhRvL..90u1601M}:
\begin{equation}\label{eq:dispersion}
  E^2\simeq p^2c^2\left(1- s \frac{2\eta}{E_{\rm Pl}} pc\right)\;,
\end{equation}
where $E$ and $p$ are the energy and momentum of a photon, $c$ is the speed of light, $\eta$ is a dimensionless
parameter describing the broken degree of Lorentz invariance, and $s=\pm1$\footnote{For vacuum dispersion
studies, an energy-dependent speed of light is predicted, and $s=+1$ or $s=-1$ corresponds to the ``subluminal'' or
``superluminal'' scenarios. Whereas, for a helicity-dependent speed of light considered here, both $s=+1$ and $s=-1$
are permissible.} represents the helicity factor.
If $s$ can take both $+1$ and $-1$ values, then Equation~(\ref{eq:dispersion}) with $\eta\neq0$ implies that
group velocities of photons with identical energy $E$ but opposite helicities (i.e., right- and left-handed
circular polarization states of the photon) should differ slightly. Consequently, the polarization vector
of a linearly polarized light may experience an energy-dependent rotation, also known as vacuum birefringence.
Observations of linear polarization can therefore be used to test Lorentz invariance (e.g.,
\citealt{1990PhRvD..41.1231C,
1998PhRvD..58k6002C,
2001PhRvD..64h3007G,
2001PhRvL..87y1304K,
2006PhRvL..97n0401K,
2007PhRvL..99a1601K,
2013PhRvL.110t1601K,
2003Natur.426Q.139M,
2004PhRvL..93b1101J,
2007MNRAS.376.1857F,
2009JCAP...08..021G,
2011PhRvD..83l1301L,
2011APh....35...95S,
2012PhRvL.109x1104T,
2013MNRAS.431.3550G,
2014MNRAS.444.2776G,
2016MNRAS.463..375L,
2017PhRvD..95h3013K,
2019PhRvD..99c5045F,
2019MNRAS.485.2401W,
2020PhRvD.102d3008F,
2020EPJP..135..527W,
2021Galax...9...44Z}).
Although the effects of vacuum birefringence are supposed to be minuscule at attainable energies,
they can increase with energy and accumulate over vast propagation distances, leading to a measurable
rotation of the plane of linear polarization as a function of energy. The rotation angle during
the propagation from the redshift $z$ to the observer is given by \citep{2011PhRvD..83l1301L,2012PhRvL.109x1104T}:
\begin{equation}\label{eq:theta-LIV}
  \Delta\phi_{\rm LIV}(E)\simeq\eta\frac{E^2}{\hbar E_{\rm Pl}}\int_0^z\frac{1+z'}{H(z')}dz'\;,
\end{equation}
where $E$ is the observed photon energy. Also, $H(z)=H_0\sqrt{\Omega_{\rm m}(1+z)^3+\Omega_{\Lambda}}$
is the cosmic expansion rate, assuming a flat $\Lambda$CDM model with Hubble constant
$H_{0}=67.4$ km $\rm s^{-1}$ $\rm Mpc^{-1}$, matter energy density $\Omega_{\rm m}=0.315$,
and vacuum energy density $\Omega_{\Lambda}=1-\Omega_{\rm m}$ \citep{2020AA...641A...6P}.

To date, the most stringent limits on the birefringent parameter $\eta$ have been obtained using
the single multi-hundred keV polarimetry of gamma-ray bursts (GRBs), yielding $\eta<\mathcal{O}(10^{-15}-10^{-16})$
\citep{2012PhRvL.109x1104T,2013MNRAS.431.3550G,2014MNRAS.444.2776G,2016MNRAS.463..375L,2019MNRAS.485.2401W}.
These upper limits stem from the argument that vacuum birefringence would produce opposite oriented
polarization vectors, thereby washing out most of the net polarization of the signal. Hence,
the detection of highly polarized sources can place upper bounds on $\eta$. Even though
such LIV tests have reached an extremely high accuracy
\citep{2012PhRvL.109x1104T,2013MNRAS.431.3550G,2014MNRAS.444.2776G,2016MNRAS.463..375L,2019MNRAS.485.2401W},
the outcomes of these upper limits on $\eta$ are lack of significantly statistical robustness.

Rather than relying on a single $\gamma$-ray polarization detection, some studies have directly
searched for an energy-dependent change in the linear polarization
angle, attributed to vacuum birefringence, within the energy-resolved spectro-polarimetric data
\citep{2007MNRAS.376.1857F,2020EPJP..135..527W,2021Galax...9...44Z}. By fitting the energy-resolved
polarization data of the optical afterglows of GRB 020813 and GRB 021004, \cite{2007MNRAS.376.1857F}
obtained a limit of $-2\times10^{-7}<\eta<1.4\times10^{-7}$ at the $3\sigma$ confidence level
(see also \citealt{2020EPJP..135..527W}). \cite{2021Galax...9...44Z} applied the same treatment
to multi-band optical polarization measurements of five blazars, and showed that Lorentz invariance
passes the strict test with a similar accuracy of $\mathcal{O}(10^{-7})$.
It is obvious from Equation~(\ref{eq:theta-LIV}) that the greater sensitivity to small values of
$\eta$ can be expected from those astrophysical sources with higher-energy polarimetry and
larger cosmological distances. Compared to $\gamma$-ray polarization constraints
\citep{2012PhRvL.109x1104T,2013MNRAS.431.3550G,2014MNRAS.444.2776G,2016MNRAS.463..375L,2019MNRAS.485.2401W},
observations of optical polarization set less stringent constraints on $\eta$ as expected
\citep{2007MNRAS.376.1857F,2020EPJP..135..527W,2021Galax...9...44Z}.

Recently, \cite{2024arXiv240613755G} presented a systematic and uniform spectro-polarimetric
analysis on five bright GRBs detected by $AstroSat$ CZTI, providing the first energy-resolved polarization
measurements in the prompt $\gamma$-ray emission of GRBs. In this work, we study Lorentz-violating effects
by analyzing the energy-dependent behavior of polarization angle during the prompt phase of these five GRBs.

\section{Energy-resolved Polarization Measurements of Prompt GRB Emission}\label{sec:data}

\cite{2024arXiv240613755G} performed an energy-resolved polarization analysis on the prompt
$\gamma$-ray emission of five bright GRBs using archival data from $AstroSat$ CZTI. The linear polarization
measurements obtained using the energy sliding binning algorithm are displayed in Figure 8 of
\cite{2024arXiv240613755G}, from which we can extract the calculated values of the energy-resolved
polarization angles and the corresponding energy segments of all the five bursts.

In order to obtain bounds on LIV with observations of $\gamma$-ray linear polarization, we also need to
know the source distances. But except for GRB 160623A at $z=0.367$ \citep{2016GCN.19708....1M}, the other
four GRBs have no measured redshifts. The well-known luminosity relation
\citep{2003MNRAS.345..743W,2004ApJ...609..935Y,2012MNRAS.421.1256N},
$\log_{10}[E_{p}(1+z)/\mathrm{keV}]=(-22.98\pm1.81)+(0.49\pm0.03)\log_{10}[L_{p}/\mathrm{erg\,s^{-1}}]$
with a standard deviation $\sigma_{\rm sc}=0.30$, is therefore adopted to estimate the redshifts of the four GRBs.
Here, $E_{p}$ is the peak energy of the burst average spectrum in the observer frame, and $L_{p}$
is the peak luminosity integrated over 1-second time intervals at the peak. We use the observed 1-second peak photon flux
and $E_{p}$ of the four bursts
[GRB 160325A: 10--1000 keV flux $8.5\;\mathrm{photons\,cm^{-2}\,s^{-1}}$ and $E_{p}=235\,\mathrm{keV}$
\citep{2016GCN.19224....1R};
GRB 160703A: 15--150 keV flux $5.8\;\mathrm{photons\,cm^{-2}\,s^{-1}}$ and $E_{p}=332.46\,\mathrm{keV}$
\citep{2016GCN.19648....1L,2024arXiv240613755G};
GRB 160802A: 10--1000 keV flux $72.5\;\mathrm{photons\,cm^{-2}\,s^{-1}}$ and $E_{p}=284\,\mathrm{keV}$
\citep{2016GCN.19754....1B};
GRB 160821A: 10--1000 keV flux $123.1\;\mathrm{photons\,cm^{-2}\,s^{-1}}$ and $E_{p}=968\,\mathrm{keV}$
\citep{2016GCN.19835....1S}] to calculate $L_{p}$ for different redshifts.
By requiring that the bursts enter the $3\sigma$ region of the luminosity relation,
we derive $z\ge0.128$ for GRB 160325A, $z\ge0.201$ for GRB 160703A, $z\ge0.052$ for GRB 160802A, and $z\ge0.108$
for GRB 160821A. Hereafter, we conservatively take the lower limits of redshifts for robust discussions on LIV.

\section{New Precision Limits on LIV}\label{sec:results}

Considering the energy-dependent rotation angle of the linear polarization plane induced by
the vacuum birefringent effect ($\Delta\phi_{\rm LIV}$; i.e., Equation~(\ref{eq:theta-LIV})),
the observed linear polarization angle ($\phi_{\rm obs}$) for photons emitted at a certain energy range
from a given astrophysical source should consist of two terms
\begin{equation}\label{eq:phi_obs}
\phi_{\rm obs}=\phi_{0}+\Delta\phi_{\rm LIV}\left(E\right)\;,
\end{equation}
where $\phi_{0}$ denotes the intrinsic polarization angle. In practice, distinguishing
the LIV-induced rotation angle $\Delta\phi_{\rm LIV}$ from an unknown intrinsic polarization angle
$\phi_{0}$ caused by the source's unknown emission mechanism is challenging. Several emission mechanisms,
such as synchrotron emission, suggested for GRB prompt emission can produce linear polarization degree
up to $\sim60\%$. In these emission mechanisms, the observed polarization angle does not depend on
photon energy \citep{1992MNRAS.258P..41R,1999PhR...314..575P,2006NJPh....8..131L,2009ApJ...698.1042T,2012ApJ...758L...1Y}.
Following \cite{2007MNRAS.376.1857F},
we simply assume that all photons in different energy channels are emitted with the same (unknown)
intrinsic polarization angle (see also \citealt{2020EPJP..135..527W,2021Galax...9...44Z}).
This assumption is susceptible to systematic uncertainties arising from the unknown intrinsic properties
of different sources. Ideally, these uncertainties can be reduced by analyzing extensive energy-resolved
polarization data for each source. However, the current observations do not provide sufficient spectro-polarimetric
data. In this study, we investigate the implications and constraints of LIV that can be established under the simplest
assumption that $\phi_{0}$ is an unknown constant. Potential evidence for vacuum birefringence (or robust limits on
the birefringent parameter $\eta$ and $\phi_{0}$) can then be directly obtained by fitting the energy-dependent
behavior of observed polarization angles with Equation~(\ref{eq:phi_obs}). The energy-resolved polarization
measurements of the prompt $\gamma$-ray emission of five GRBs \citep{2024arXiv240613755G} are utilized in our analysis.

\begin{figure*}
\includegraphics[width=0.5\textwidth]{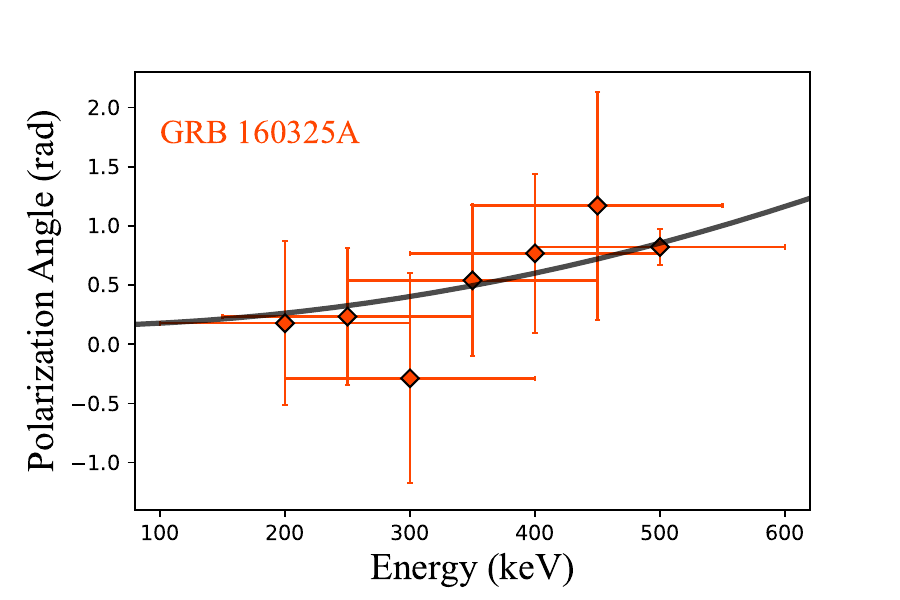}
\includegraphics[width=0.5\textwidth]{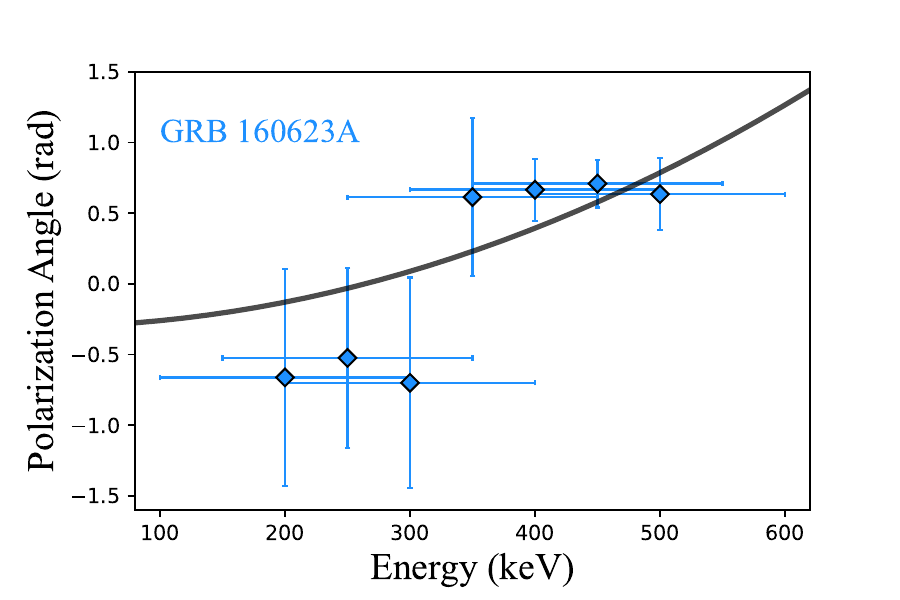}
\includegraphics[width=0.5\textwidth]{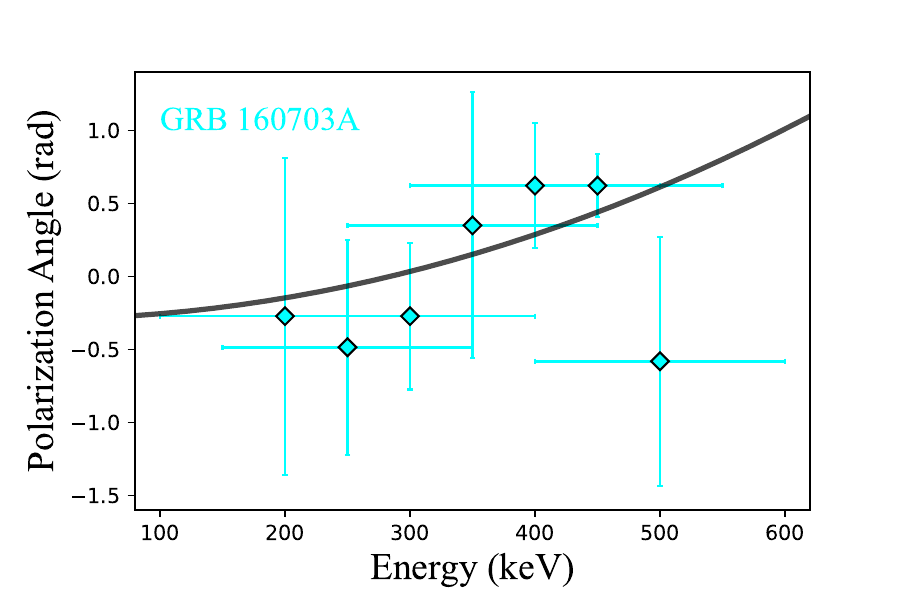}
\includegraphics[width=0.5\textwidth]{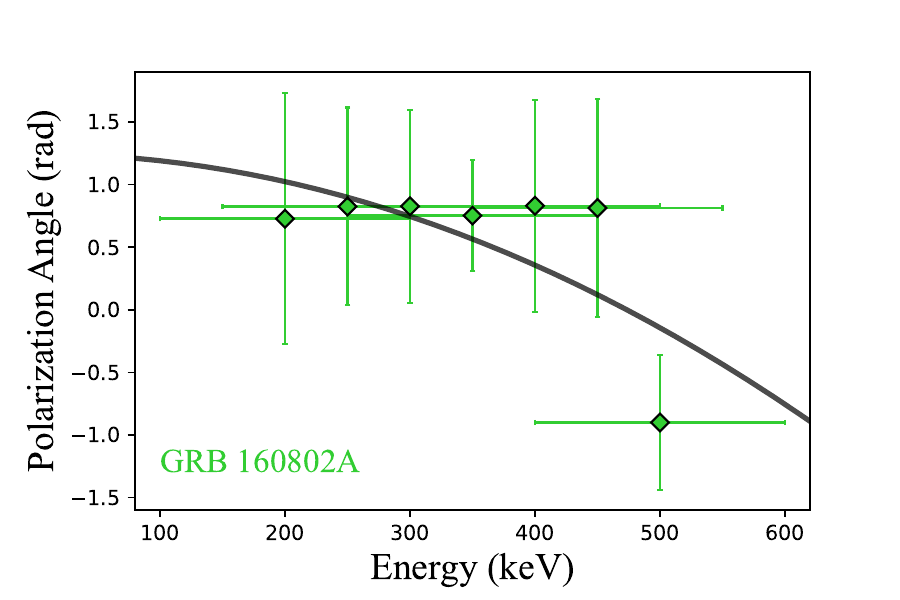}
\includegraphics[width=0.5\textwidth]{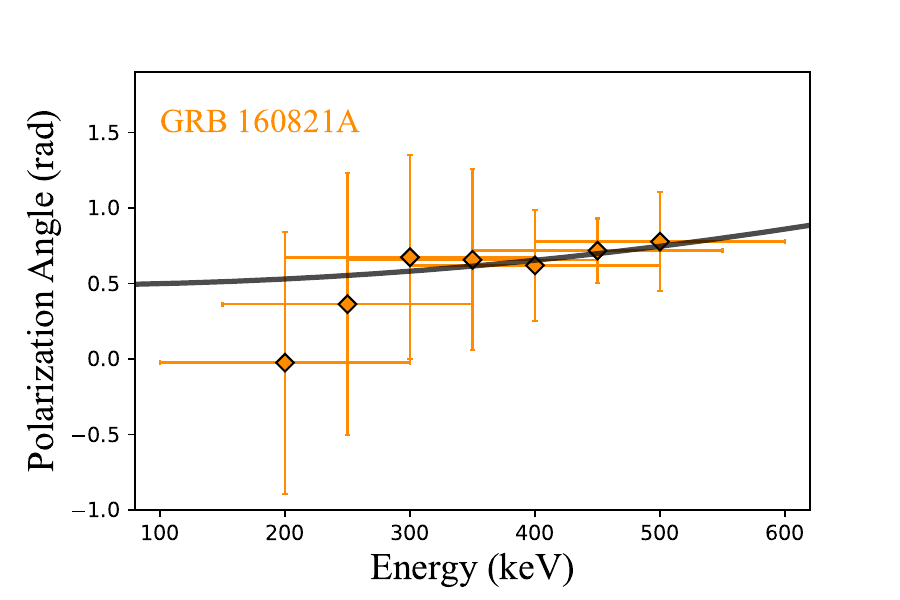}
\caption{The polarization angles as a function of energy. Data points are the energy-resolved
polarization measurements of five GRBs, and the vertical and horizontal error bars, respectively,
correspond to the measurement errors of the polarization angles and the ranges of each energy bin.
Curves show the optimal predictions of the LIV model (see text and Equation~(\ref{eq:phi_obs})).}
\label{fig1}
\end{figure*}

The observed polarization angles as a function of energy for all five GRBs are shown in Figure~\ref{fig1}.
In principle, the two free parameters ($\eta$ and $\phi_{0}$) can be optimized by directly fitting the energy-resolved
polarization data of each GRB. However, the polarization measurements reported in \cite{2024arXiv240613755G}
are not statistically independent due to overlapping energy bins, as indicated by the error bars on the X-axis
in Figure~\ref{fig1}. Therefore, the covariance matrix for the energy bins should be considered when using
these polarization data for parameter estimation. To address this, we give a statistical method to derive
the covariance matrix. The key steps of this process are as follows:
\begin{enumerate}
    \item  Generate 10,000 realizations of a sample resembling the data by drawing $n$ energy measurements
    using the Gaussian distribution $\mathcal{N}(E,\,\sigma_{E})$, where $n$ is the number of energy bins
    in our polarization data. Here, $E$ and $\sigma_{E}=(E_{\rm max}-E_{\rm min})/2$ are the median value
    and the error of the corresponding energy bin, with $E_{\rm min}$ and $E_{\rm max}$ being the left and
    right edges of each energy bin.

    \item For any two energy bins with median values $E_i$ and $E_j$, the covariance between them can be calculated
    by comparing 10,000 simulated $E_i$ values and 10,000 simulated $E_j$ values using
    \begin{equation}\label{eq:cov_E}
      \mathrm{\bf Cov}_{E}(E_{i},\,E_{j}) = \frac{1}{N-1}\sum_{k=1}^{N}\left[ \left(E_{ik}-\bar{E_{i}}\right) \left(E_{jk}-\bar{E_{j}}\right) \right]\;,
    \end{equation}
    where $N=10,000$ is the number of simulated $E$ data points, and $\bar{E}$ is the average value of
    the 10,000 simulated $E$ data points.
\end{enumerate}

Now, the parameters $\eta$ and $\phi_{0}$ for each GRB can be optimized using
the Bayesian inference method by defining the likelihood function:
\begin{equation}\label{eq:likelihood}
 \mathcal{L} = \frac{\exp{\left[-\frac{1}{2}(\hat{\mathbf{\phi}}_{\rm obs}-\hat{\mathbf{\phi}}_{\rm th})^{T}
 {\bf Cov}^{-1}(\hat{\mathbf{\phi}}_{\rm obs}-\hat{\mathbf{\phi}}_{\rm th})
 \right]}}{\sqrt{{\left(2\pi\right)^n}\det{\bf Cov}}}\;,
\end{equation}
where $\hat{\mathbf{\phi}}_{\rm obs}(\bf{\emph{E}})$ and $\hat{\mathbf{\phi}}_{\rm th}(\eta,\,\phi_{0};\,\bf{\emph{E}})$
are the observed and theoretical (see Equation~(\ref{eq:phi_obs})) polarization angle vectors with $n$ components,
$n$ being the number of polarization angles, and $\mathrm{\bf Cov}$ is the full $n\times n$ covariance matrix, defined by
\begin{equation}
{\bf Cov}=\mathrm{diag}\left(\sigma_{\phi_{\rm obs}}^{2}\right)+{\bf Cov}_{\phi}\;.
\end{equation}
In this context, $\mathrm{diag}(\sigma_{\phi_{\rm obs}}^{2})$ represents the diagonal component of the measurement variance
in $\phi_{\rm obs}$, and ${\bf Cov}_{\phi}$ is the covariance matrix propagated from that of the energy bins
$\mathrm{\bf Cov}_{E}$, given by
\begin{equation}
{\bf Cov}_{\phi}(\eta;\,E_{i},\,E_{j})=\mathcal{C}_{i}\mathcal{C}_{j}{\bf Cov}_{E}(E_{i},\,E_{j})\;,
\end{equation}
where the coefficient $\mathcal{C}_{i}$ in error propagation is a function of $\eta$,
\begin{equation}
\mathcal{C}_{i}=2\frac{\Delta\phi_{\rm LIV}\left(\eta;\,E_{i}\right)}{E_{i}}\sigma_{E_i}\;.
\end{equation}

\begin{figure}
\begin{center}
\includegraphics[width=0.45\textwidth]{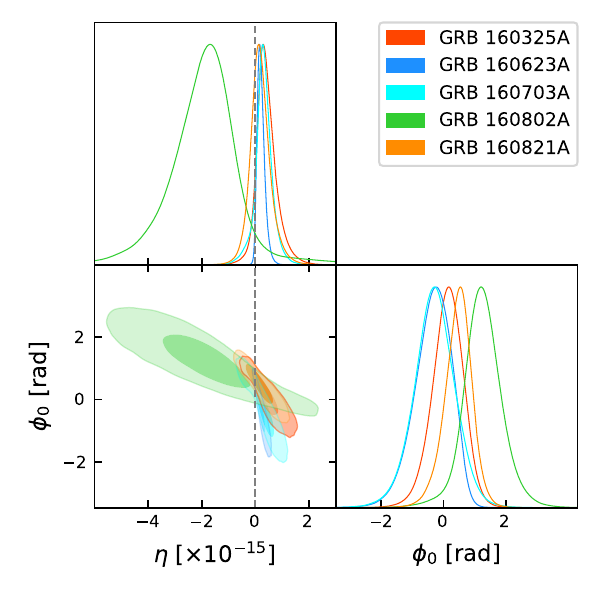}
\vskip-0.1in
\caption{1D marginalized probability distributions and 2D contours with 1--2$\sigma$ confidence levels
for the birefringent parameter $\eta$ and the intrinsic polarization angle $\phi_{0}$,
constrained by the energy-resolved polarization data of the five different GRBs. The vertical
dashed line corresponds to the case of no LIV (i.e., $\eta=0$).}
\label{fig2}
\vskip-0.2in
\end{center}
\end{figure}

It should be underlined that each theoretical polarization angle $\phi_{{\rm th},i}$ depends on
$\eta$ and $\phi_{0}$. We maximize the likelihood $\mathcal{L}$ over the two free parameters.
The Python Markov chain Monte Carlo module, EMCEE \citep{2013PASP..125..306F}, is applied to explore
the posterior probability distributions of the free parameters. The 1D marginalized probability distributions
and 2D contours with 1--2$\sigma$ confidence levels for the two parameters, constrained by the energy-resolved
polarization data of each GRB, are presented in Figure~\ref{fig2}. These contours show that at the 95\%
confidence level, the inferred values are $\eta=(0.4^{+1.0}_{-0.7})\times10^{-15}$ and
$\phi_{0}=0.15^{+0.94}_{-1.11}\,\mathrm{rad}$ for GRB 160325A,
$\eta=(0.2^{+0.4}_{-0.2})\times10^{-15}$ and
$\phi_{0}=-0.30^{+0.96}_{-1.36}\,\mathrm{rad}$ for GRB 160623A,
$\eta=(0.3^{+0.7}_{-0.8})\times10^{-15}$ and
$\phi_{0}=-0.29^{+1.32}_{-1.39}\,\mathrm{rad}$ for GRB 160703A,
$\eta=(-1.9^{+3.1}_{-3.0})\times10^{-15}$ and
$\phi_{0}=1.25^{+1.32}_{-1.33}\,\mathrm{rad}$ for GRB 160802A, and
$\eta=(0.2^{+1.0}_{-0.7})\times10^{-15}$ and
$\phi_{0}=0.49^{+0.82}_{-1.02}\,\mathrm{rad}$ for GRB 160821A.
The resulting constraints on $\eta$ and $\phi_{0}$ for each GRB data are summarized in Table~\ref{tab1}.
We can see that all the inferred values of $\eta$ are consistent with 0 at the $2\sigma$ confidence level,
implying that there is no evidence of LIV. To illustrate the fits, the energy-dependence of polarization angle
expected from the LIV model (see Equation~(\ref{eq:phi_obs}); with each set of the best-fit parameters) are shown
as solid curves in Figure~\ref{fig1}.

\begin{table}
\renewcommand\arraystretch{1.3}
\tabcolsep=0.4cm
\centering \caption{Best-fit Results (with 95\% Confidence-level Uncertainties) on
the Birefringent Parameter $\eta$ and the Intrinsic Polarization Angle $\phi_{0}$
from Energy-resolved Polarization Measurements of Five GRBs}
\begin{tabular}{CCCC}
\hline
\hline
\mathrm{Source}	&  $z$	&  $\eta\,(\times 10^{-15})$   &   $\phi_{0}\,(\mathrm{rad})$\\
\hline
\mathrm{GRB 160325A}		&   0.128\textsuperscript{a}		&   $0.4^{+1.0}_{-0.7}$   &  $0.15^{+0.94}_{-1.11}$ \\
\mathrm{GRB 160623A}		&   0.367		                    &   $0.2^{+0.4}_{-0.2}$   &  $-0.30^{+0.96}_{-1.36}$ \\
\mathrm{GRB 160703A}	    &   0.201\textsuperscript{a}		&   $0.3^{+0.7}_{-0.8}$   &  $-0.29^{+1.32}_{-1.39}$ \\
\mathrm{GRB 160802A}		&   0.052\textsuperscript{a}		&   $-1.9^{+3.1}_{-3.0}$   &  $1.25^{+1.32}_{-1.33}$ \\
\mathrm{GRB 160821A}		&   0.108\textsuperscript{a}		&   $0.2^{+1.0}_{-0.7}$   &  $0.49^{+0.82}_{-1.02}$ \\
\hline
\end{tabular}
\label{tab1}
\noindent{\footnotesize{\textsuperscript{a} The redshifts of the four GRBs are estimated by the luminosity relation.}}
\end{table}

Compared with previous results obtained from multi-band optical polarization measurements
($|\eta|<2\times10^{-7}$; \citealt{2007MNRAS.376.1857F,2020EPJP..135..527W,2021Galax...9...44Z}),
our limits on $\eta$ represent an improvement of at least eight orders of magnitude.
While our limits are essentially as good as previous best bounds from $\gamma$-ray
polarimetry of other GRBs ($\eta<\mathcal{O}(10^{-15}-10^{-16})$;
\citealt{2012PhRvL.109x1104T,2013MNRAS.431.3550G,2014MNRAS.444.2776G,2016MNRAS.463..375L,2019MNRAS.485.2401W}),
there is merit to the results. Thanks to the adoption of the energy-resolved polarization data,
our constraints on $\eta$ could be statistically more robust compared to previous results,
which were based on a single polarization measurement in the 100s keV energy range.

As mentioned in \cite{2012PhRvL.109x1104T}, if the rotation angle differs by more than $\pi/2$
over a broad bandwidth, the net polarization of the signal would be severely suppressed and
would not reach the observed level. If the suppression is strong, the polarization signal
will become so weak that the energy-dependent rotation of the linear polarization plane, as discussed in this
paper, may not be measurable. However, the effects of vacuum birefringence can amplify
with increasing energy (see Equation~(\ref{eq:theta-LIV})), resulting in a measurable rotation.
Theoretically, for a source at $z\sim1$, the rotation angle is given by $\Delta\phi_{\rm LIV}=6.3\times10^{10}
\eta(E/\mathrm{keV})^2\,\mathrm{rad}$. Therefore, the relevant energies for observing
this birefringent effect are $E_{\rm obs} \geq E_{\rm bir}\equiv4.0\times10^{-6}\eta^{-1/2}\,\mathrm{keV}$,
where $E_{\rm bir}$ is defined as $\Delta\phi_{\rm LIV}(E_{\rm bir})=1\,\mathrm{rad}$. This implies that even with
a very small $\eta \sim 10^{-16}$, the birefringent effect can be detected through polarimetry of
$\sim400\,\mathrm{keV}$ photons from a source at $z\sim1$. In other words, as long as the intrinsic polarization
angle $\phi_{0}$ is well known, the LIV-induced rotation angle $\Delta\phi_{\rm LIV}$ is measurable
using $\gamma$-ray polarimeters.

We note that the polarization angles of GRB 160325A, GRB 160623A, GRB 160703A, and GRB 160802A
measured at different time intervals remain consistent within their respective error bars.
In contrast, the time-resolved polarization analysis of GRB 160821A exhibits temporal variations
in the polarization angle \citep{2019ApJ...882L..10S,2024arXiv240613755G}. From theoretical calculations
(e.g., \citealt{2021Galax...9...82G}) and observations (e.g., \citealt{2019ApJ...882L..10S,2024arXiv240613755G}),
the polarization angle has been observed to change with time. Strictly speaking,
to constrain LIV, we should allow for variations in spectro-polarimetry across different
time intervals and consider the average spectro-polarimetric angle over time for analysis.
However, the energy-resolved polarization measurements of GRB 160821A were only carried out
within a specific temporal window \citep{2024arXiv240613755G}. Fortunately,
the resulting constraints on the birefringent parameter $\eta$ mainly depend on the working
energy band of the polarimeter and the redshift of the source (see Equation~(\ref{eq:theta-LIV})).
With the same source and polarimeter, the temporal variations in the spectro-polarimetric angles
have minimal impact on our results.

\section{Summary and Discussions}\label{sec:conclusions}
Violations of Lorentz invariance can lead to vacuum birefringence of light, which results in an energy-dependent
rotation of the polarization plane of linearly polarized photons. Lorentz invariance can therefore be tested with
astrophysical polarization measurements. Very recently, \cite{2024arXiv240613755G} reported
the energy-resolved polarization measurements in the prompt $\gamma$-ray emission of five bright GRBs detected by
$AstroSat$ CZTI. In this work, we investigated the implications and limits on LIV that can be set from these unique
polarization observations.

Assuming an unknown constant for the intrinsic polarization angle, we searched for the energy-dependent change of
the linear polarization angle, resulting from the birefringent effect, in the spectro-polarimetric data of
these five GRBs. By fitting the polarization angle and energy measurements of each GRB, we place a statistically
robust limit on the birefringent parameter $\eta$ quantifying the broken degree of Lorentz invariance.
For instance, with the data of GRB 160623A, we have $-0.1\times10^{-16}<\eta<5.8\times10^{-16}$ at the $2\sigma$
confidence level. Similar $\eta$ constraints have been obtained from the other four GRBs.
Our new results represent sensitivities improved by at least eight orders of magnitude over existing $\eta$
bounds from multi-band optical polarization measurements. Moreover, our constraints are competitive with
previous best bounds from $\gamma$-ray polarimetry of other GRBs. Unlike previous analyses that only
rely on a single polarization measurement in the 100s keV range, our constraints derived from energy-resolved
polarimetric data set are statistically more robust.

There are some other astrophysical effects that can also produce an energy-dependent rotation
of the linear polarization plane, such as Faraday rotation and the violation of the weak equivalence principle (WEP).

\begin{enumerate}
    \item It is well known that magnetized plasmas can affect the rotation of the linear polarization plane
(the so-called Faraday rotation). The dependence of the rotation angle on Faraday rotation is
$\Delta\phi_{\rm Far}\propto E^{-2}$, quite different from its dependence on LIV effects as
$\Delta\phi_\mathrm{LIV}\propto E^{2}$ shown in Equation~(\ref{eq:theta-LIV}). The Faraday rotation
angle $\Delta\phi_{\rm Far}$ would be significant at the low-frequency radio band. But at optical and
higher-energy bands, such as the $\gamma$-ray energies considered here, $\Delta\phi_{\rm Far}$ is negligible.

    \item A possible violation of the WEP can also produce vacuum birefringence,
    which leads to an energy-dependent rotation of the linear polarization plane \citep{2017MNRAS.469L..36Y,2019PhRvD..99j3012W,2020EPJP..135..527W}.
    The rotation angle induced by the WEP violation is
    $\Delta\phi_{\rm WEP}=\left|\frac{\Delta \gamma}{c^3}\int_{r_e}^{r_o}~U(r)dr\right| \frac{E}{\hbar}$,
    where $\Delta \gamma$ is the difference of the parameterized post-Newtonian parameter
    $\gamma$ values describing the deviation from the WEP, and $U(r)$ represents the
    gravitational potential, which is sourced by the matter density along the propagation path
    from the source $r_e$ to the observer $r_o$. \cite{2019PhRvD..99j3012W} has provided an upper
    limit for the $\gamma$ discrepancy as $\Delta \gamma < \mathcal{O}(10^{-33})$. In this case,
    $\Delta\phi_{\rm WEP}$ is negligible in our analysis.
\end{enumerate}

Besides polarization measurements, vacuum dispersion time-of-flight measurements from
astrophysical sources have also been widely used to test Lorentz invariance (e.g.,
\citealt{1998Natur.393..763A,2013PhRvD..87l2001V,2024PhRvL.133g1501C}). Using the time-of-flight measurements
of the TeV emission from the brightest-of-all-time GRB 221009A, \cite{2024PhRvL.133g1501C}
set the current best limit on the QG energy scale, which is the energy scale where
LIV-induced QG effects become significant ($E_{\rm QG}$), as $E_{\rm QG}>1.0\times10^{20}\,\mathrm{GeV}$.
Note that there is a relation between the birefringent parameter $\eta$ and $E_{\rm QG}$,
$E_{\rm QG}=E_\mathrm{Pl}/2\eta$ \citep{Wei2022}. It is thus interesting to make a comparison of
the latest achievements in sensitivity of polarization measurements versus time-of-flight measurements.
In this work, the energy-resolved polarimetry of GRB 160623A yielded the strictest upper limit on
the birefringent parameter, $\eta<5.8\times10^{-16}$, which corresponds to $E_{\rm QG}>1.1\times10^{34}\,\mathrm{GeV}$.
Obviously, this polarization constraint is about 14 orders of magnitude stronger than the time-of-flight constraint.
Future polarization measurements of astrophysical sources such as GRBs at higher $\gamma$-ray energies
and larger distances would further enhance sensitivity to LIV tests through the vacuum birefringent effect.

\begin{acknowledgments}
We are grateful to the anonymous referee for the valuable comments, which have significantly improved
the presentation of the paper. We also thank Bao Wang for his helpful discussions on the covariance matrix.
This work is supported by the Strategic Priority Research Program of the Chinese Academy
of Sciences (grant No. XDB0550400), the National Key R\&D Program of China (2024YFA1611704),
the National Natural Science Foundation of China (grant Nos. 12422307, 12373053, and 12321003),
the Key Research Program of Frontier Sciences (grant No. ZDBS-LY-7014) of Chinese Academy of
Sciences, and the Natural Science Foundation of Jiangsu Province (grant No. BK20221562).
\end{acknowledgments}


\begin{thebibliography}{}
\expandafter\ifx\csname natexlab\endcsname\relax\def\natexlab#1{#1}\fi
\providecommand{\url}[1]{\href{#1}{#1}}
\providecommand{\dodoi}[1]{doi:~\href{http://doi.org/#1}{\nolinkurl{#1}}}
\providecommand{\doeprint}[1]{\href{http://ascl.net/#1}{\nolinkurl{http://ascl.net/#1}}}
\providecommand{\doarXiv}[1]{\href{https://arxiv.org/abs/#1}{\nolinkurl{https://arxiv.org/abs/#1}}}

\bibitem[{{Addazi} {et~al.}(2022){Addazi}, {Alvarez-Muniz}, {Alves Batista},
  {Amelino-Camelia}, {Antonelli}, {Arzano}, {Asorey}, {Atteia}, {Bahamonde},
  {Bajardi}, {Ballesteros}, {Baret}, {Barreiros}, {Basilakos}, {Benisty},
  {Birnholtz}, {Blanco-Pillado}, {Blas}, {Bolmont}, {Boncioli}, {Bosso},
  {Calcagni}, {Capozziello}, {Carmona}, {Cerci}, {Chernyakova}, {Clesse},
  {Coelho}, {Colak}, {Cortes}, {Das}, {D'Esposito}, {Demirci}, {Di Luca}, {di
  Matteo}, {Dimitrijevic}, {Djordjevic}, {Prester}, {Eichhorn}, {Ellis},
  {Escamilla-Rivera}, {Fabiano}, {Franchino-Vi{\~n}as}, {Frassino},
  {Frattulillo}, {Funk}, {Fuster}, {Gamboa}, {Gent}, {Gergely}, {Giammarchi},
  {Giesel}, {Glicenstein}, {Gracia-Bond{\'\i}a}, {Gracia-Ruiz}, {Gubitosi},
  {Guendelman}, {Gutierrez-Sagredo}, {Haegel}, {Heefer}, {Held}, {Herranz},
  {Hinderer}, {Illana}, {Ioannisian}, {Jetzer}, {Joaquim}, {Kampert}, {Uysal},
  {Katori}, {Kazarian}, {Kerszberg}, {Kowalski-Glikman}, {Kuroyanagi},
  {L{\"a}mmerzahl}, {Said}, {Liberati}, {Lim}, {Lobo}, {L{\'o}pez-Moya},
  {Luciano}, {Manganaro}, {Marcian{\`o}}, {Mart{\'\i}n-Moruno}, {Martinez},
  {Martinez}, {Mart{\'\i}nez-Huerta}, {Mart{\'\i}nez-Mirav{\'e}}, {Masip},
  {Mattingly}, {Mavromatos}, {Mazumdar}, {M{\'e}ndez}, {Mercati}, {Micanovic},
  {Mielczarek}, {Miller}, {Milosevic}, {Minic}, {Miramonti}, {Mitsou}, {Moniz},
  {Mukherjee}, {Nardini}, {Navas}, {Niechciol}, {Nielsen}, {Obers},
  {Oikonomou}, {Oriti}, {Paganini}, {Palomares-Ruiz}, {Pasechnik}, {Pasic},
  {P{\'e}rez de los Heros}, {Pfeifer}, {Pieroni}, {Piran}, {Platania},
  {Rastgoo}, {Relancio}, {Reyes}, {Ricciardone}, {Risse}, {Frias}, {Rosati},
  {Rubiera-Garcia}, {Sahlmann}, {Sakellariadou}, {Salamida}, {Saridakis},
  {Satunin}, {Schiffer}, {Sch{\"u}ssler}, {Sigl}, {Sitarek}, {Peracaula},
  {Sopuerta}, {Sotiriou}, {Spurio}, {Staicova}, {Stergioulas}, {Stoica},
  {Stri{\v{s}}kovi{\'c}}, {Stuttard}, {Cerci}, {Tavakoli}, {Ternes},
  {Terzi{\'c}}, {Thiemann}, {Tinyakov}, {Torri}, {T{\'o}rtola}, {Trimarelli},
  {Trze{\'s}niewski}, {Tureanu}, {Urban}, {Vagenas}, {Vernieri}, {Vitagliano},
  {Wallet}, \& {Zornoza}}]{2022PrPNP.12503948A}
{Addazi}, A., {Alvarez-Muniz}, J., {Alves Batista}, R., {et~al.} 2022, Progress
  in Particle and Nuclear Physics, 125, 103948,
  \dodoi{10.1016/j.ppnp.2022.103948}

\bibitem[{{Alfaro} {et~al.}(2002){Alfaro}, {Morales-T{\'e}cotl}, \&
  {Urrutia}}]{2002PhRvD..65j3509A}
{Alfaro}, J., {Morales-T{\'e}cotl}, H.~A., \& {Urrutia}, L.~F. 2002, \prd, 65,
  103509, \dodoi{10.1103/PhysRevD.65.103509}

\bibitem[{{Alves Batista} {et~al.}(2023){Alves Batista}, {Amelino-Camelia},
  {Boncioli}, {Carmona}, {di Matteo}, {Gubitosi}, {Lobo}, {Mavromatos},
  {Pfeifer}, {Rubiera-Garcia}, {Saridakis}, {Terzi{\'c}}, {Vagenas}, {Vargas
  Moniz}, {Abdalla}, {Adamo}, {Addazi}, {Anagnostopoulos}, {Antonelli},
  {Asorey}, {Ballesteros}, {Basilakos}, {Benisty}, {Boettcher}, {Bolmont},
  {Bonilla}, {Bosso}, {Bouhmadi-L{\'o}pez}, {Burderi}, {Campoy-Ordaz},
  {Caroff}, {Cerci}, {Cortes}, {D'Esposito}, {Das}, {de Cesare}, {Demirci}, {Di
  Lodovico}, {Di Salvo}, {Diego}, {Djordjevic}, {Domi}, {Ducobu},
  {Escamilla-Rivera}, {Fabiano}, {Fern{\'a}ndez-Silvestre},
  {Franchino-Vi{\~n}as}, {Frassino}, {Frattulillo}, {Garay}, {Gaug}, {Gergely},
  {Guendelman}, {Guetta}, {Gutierrez-Sagredo}, {He}, {Heefer}, {Juri{\'c}},
  {Katori}, {Kowalski-Glikman}, {Lambiase}, {Levi Said}, {Li}, {Li}, {Luciano},
  {Ma}, {Marciano}, {Martinez}, {Mazumdar}, {Menezes}, {Mercati}, {Minic},
  {Miramonti}, {Mitsou}, {Mustamin}, {Navas}, {Olmo}, {Oriti}, {{\"O}vg{\"u}n},
  {Pantig}, {Parvizi}, {Pasechnik}, {Pasic}, {Petruzziello}, {Platania},
  {Rasouli}, {Rastgoo}, {Relancio}, {Rescic}, {Reyes}, {Rosati}, {Sakall{\i}},
  {Salamida}, {Sanna}, {Staicova}, {Stri{\v{s}}kovi{\'c}}, {Sunar Cerci},
  {Torri}, {Vigliano}, {Wagner}, {Wallet}, {Wojnar}, {Zarikas}, {Zhu}, \&
  {Zornoza}}]{2023arXiv231200409A}
{Alves Batista}, R., {Amelino-Camelia}, G., {Boncioli}, D., {et~al.} 2023,
  arXiv e-prints, arXiv:2312.00409, \dodoi{10.48550/arXiv.2312.00409}

\bibitem[{{Amelino-Camelia}(2002)}]{2002Natur.418...34A}
{Amelino-Camelia}, G. 2002, \nat, 418, 34, \dodoi{10.1038/418034a}

\bibitem[{{Amelino-Camelia}(2013)}]{2013LRR....16....5A}
---. 2013, Living Reviews in Relativity, 16, 5, \dodoi{10.12942/lrr-2013-5}

\bibitem[{{Amelino-Camelia} {et~al.}(1997){Amelino-Camelia}, {Ellis},
  {Mavromatos}, \& {Nanopoulos}}]{1997IJMPA..12..607A}
{Amelino-Camelia}, G., {Ellis}, J., {Mavromatos}, N.~E., \& {Nanopoulos}, D.~V.
  1997, International Journal of Modern Physics A, 12, 607,
  \dodoi{10.1142/S0217751X97000566}

\bibitem[{{Amelino-Camelia} {et~al.}(1998){Amelino-Camelia}, {Ellis},
  {Mavromatos}, {Nanopoulos}, \& {Sarkar}}]{1998Natur.393..763A}
{Amelino-Camelia}, G., {Ellis}, J., {Mavromatos}, N.~E., {Nanopoulos}, D.~V.,
  \& {Sarkar}, S. 1998, \nat, 393, 763, \dodoi{10.1038/31647}

\bibitem[{{Bissaldi}(2016)}]{2016GCN.19754....1B}
{Bissaldi}, E. 2016, GRB Coordinates Network, 19754, 1

\bibitem[{{Cao} {et~al.}(2024){Cao}, {Aharonian}, {Axikegu}, {Bao}, {Bastieri},
  {Bi}, {Bi}, {Bian}, {Bukevich}, {Cao}, {Cao}, {Cao}, {Chang}, {Chang},
  {Chen}, {Chen}, {Chen}, {Chen}, {Chen}, {Chen}, {Chen}, {Chen}, {Chen},
  {Chen}, {Chen}, {Chen}, {Chen}, {Chen}, {Cheng}, {Cheng}, {Cui}, {Cui},
  {Cui}, {Cui}, {Dai}, {Dai}, {Dai}, {Danzengluobu}, {Duan}, {Fan}, {Fan},
  {Fang}, {Fang}, {Fang}, {Feng}, {Feng}, {Feng}, {Feng}, {Feng}, {Feng},
  {Feng}, {Gabici}, {Gao}, {Gao}, {Gao}, {Gao}, {Gao}, {Ge}, {Geng},
  {Giacinti}, {Gong}, {Gou}, {Gu}, {Guo}, {Guo}, {Guo}, {Guo}, {Han}, {Hasan},
  {He}, {He}, {He}, {He}, {Hor}, {Hou}, {Hou}, {Hou}, {Hu}, {Hu}, {Hu},
  {Huang}, {Huang}, {Huang}, {Huang}, {Huang}, {Huang}, {Ji}, {Jia}, {Jia},
  {Jiang}, {Jiang}, {Jiang}, {Jin}, {Kang}, {Karpikov}, {Kuleshov}, {Kurinov},
  {Li}, {Li}, {Li}, {Li}, {Li}, {Li}, {Li}, {Li}, {Li}, {Li}, {Li}, {Li}, {Li},
  {Li}, {Li}, {Li}, {Li}, {Li}, {Li}, {Liang}, {Liang}, {Lin}, {Liu}, {Liu},
  {Liu}, {Liu}, {Liu}, {Liu}, {Liu}, {Liu}, {Liu}, {Liu}, {Liu}, {Liu}, {Liu},
  {Liu}, {Luo}, {Luo}, {Lv}, {Ma}, {Ma}, {Ma}, {Mao}, {Min}, {Mitthumsiri},
  {Mu}, {Nan}, {Neronov}, {Ou}, {Pattarakijwanich}, {Pei}, {Qi}, {Qi}, {Qiao},
  {Qin}, {Raza}, {Ruffolo}, {S{\'a}iz}, {Saeed}, {Semikoz}, {Shao},
  {Shchegolev}, {Sheng}, {Shu}, {Song}, {Stenkin}, {Stepanov}, {Su}, {Sun},
  {Sun}, {Sun}, {Sun}, {Takata}, {Tam}, {Tang}, {Tang}, {Tang}, {Tian}, {Wang},
  {Wang}, {Wang}, {Wang}, {Wang}, {Wang}, {Wang}, {Wang}, {Wang}, {Wang},
  {Wang}, {Wang}, {Wang}, {Wang}, {Wang}, {Wang}, {Wang}, {Wang}, {Wang},
  {Wang}, {Wang}, {Wang}, {Wei}, {Wei}, {Wei}, {Wen}, {Wu}, {Wu}, {Wu}, {Wu},
  {Wu}, {Wu}, {Xi}, {Xia}, {Xiang}, {Xiao}, {Xiao}, {Xin}, {Xing}, {Xiong},
  {Xiong}, {Xu}, {Xu}, {Xu}, {Xu}, {Xue}, {Yan}, {Yan}, {Yan}, {Yang}, {Yang},
  {Yang}, {Yang}, {Yang}, {Yang}, {Yang}, {Yang}, {Yao}, {Yao}, {Yin}, {Yin},
  {You}, {You}, {Yu}, {Yuan}, {Yue}, {Zeng}, {Zeng}, {Zeng}, {Zha}, {Zhang},
  {Zhang}, {Zhang}, {Zhang}, {Zhang}, {Zhang}, {Zhang}, {Zhang}, {Zhang},
  {Zhang}, {Zhang}, {Zhang}, {Zhang}, {Zhang}, {Zhang}, {Zhang}, {Zhang},
  {Zhang}, {Zhao}, {Zhao}, {Zhao}, {Zhao}, {Zhao}, {Zhao}, {Zheng}, {Zhong},
  {Zhou}, {Zhou}, {Zhou}, {Zhou}, {Zhou}, {Zhou}, {Zhou}, {Zhou}, {Zhu}, {Zhu},
  {Zhu}, {Zhu}, {Zhu}, {Zou}, {Zuo}, \& {Lhaaso
  Collaboration}}]{2024PhRvL.133g1501C}
{Cao}, Z., {Aharonian}, F., {Axikegu}, Bai, Y.~X., {et~al.} 2024, \prl, 133,
  071501, \dodoi{10.1103/PhysRevLett.133.071501}

\bibitem[{{Carroll} {et~al.}(1990){Carroll}, {Field}, \&
  {Jackiw}}]{1990PhRvD..41.1231C}
{Carroll}, S.~M., {Field}, G.~B., \& {Jackiw}, R. 1990, \prd, 41, 1231,
  \dodoi{10.1103/PhysRevD.41.1231}

\bibitem[{{Colladay} \& {Kosteleck{\'y}}(1998)}]{1998PhRvD..58k6002C}
{Colladay}, D., \& {Kosteleck{\'y}}, V.~A. 1998, \prd, 58, 116002,
  \dodoi{10.1103/PhysRevD.58.116002}

\bibitem[{Desai(2024)}]{Desai2024}
Desai, S. 2024, Astrophysical and Cosmological Searches for Lorentz Invariance
  Violation (Singapore: Springer Nature Singapore), 433--463,
  \dodoi{10.1007/978-981-97-2871-8_11}

\bibitem[{{Ellis} {et~al.}(1999){Ellis}, {Mavromatos}, \&
  {Nanopoulos}}]{1999GReGr..31.1257E}
{Ellis}, J., {Mavromatos}, N.~E., \& {Nanopoulos}, D.~V. 1999, General
  Relativity and Gravitation, 31, 1257, \dodoi{10.1023/A:1026720723556}

\bibitem[{{Fan} {et~al.}(2007){Fan}, {Wei}, \& {Xu}}]{2007MNRAS.376.1857F}
{Fan}, Y.-Z., {Wei}, D.-M., \& {Xu}, D. 2007, \mnras, 376, 1857,
  \dodoi{10.1111/j.1365-2966.2007.11576.x}

\bibitem[{{Foreman-Mackey} {et~al.}(2013){Foreman-Mackey}, {Hogg}, {Lang}, \&
  {Goodman}}]{2013PASP..125..306F}
{Foreman-Mackey}, D., {Hogg}, D.~W., {Lang}, D., \& {Goodman}, J. 2013, \pasp,
  125, 306, \dodoi{10.1086/670067}

\bibitem[{{Friedman} {et~al.}(2020){Friedman}, {Gerasimov}, {Leon}, {Stevens},
  {Tytler}, {Keating}, \& {Kislat}}]{2020PhRvD.102d3008F}
{Friedman}, A.~S., {Gerasimov}, R., {Leon}, D., {et~al.} 2020, \prd, 102,
  043008, \dodoi{10.1103/PhysRevD.102.043008}

\bibitem[{{Friedman} {et~al.}(2019){Friedman}, {Leon}, {Crowley}, {Johnson},
  {Teply}, {Tytler}, {Keating}, \& {Cole}}]{2019PhRvD..99c5045F}
{Friedman}, A.~S., {Leon}, D., {Crowley}, K.~D., {et~al.} 2019, \prd, 99,
  035045, \dodoi{10.1103/PhysRevD.99.035045}

\bibitem[{{Gill} {et~al.}(2021){Gill}, {Kole}, \&
  {Granot}}]{2021Galax...9...82G}
{Gill}, R., {Kole}, M., \& {Granot}, J. 2021, Galaxies, 9, 82,
  \dodoi{10.3390/galaxies9040082}

\bibitem[{{Gleiser} \& {Kozameh}(2001)}]{2001PhRvD..64h3007G}
{Gleiser}, R.~J., \& {Kozameh}, C.~N. 2001, \prd, 64, 083007,
  \dodoi{10.1103/PhysRevD.64.083007}

\bibitem[{{G{\"o}tz} {et~al.}(2013){G{\"o}tz}, {Covino}, {Fern{\'a}ndez-Soto},
  {Laurent}, \& {Bo{\v s}njak}}]{2013MNRAS.431.3550G}
{G{\"o}tz}, D., {Covino}, S., {Fern{\'a}ndez-Soto}, A., {Laurent}, P., \&
  {Bo{\v s}njak}, {\v Z}. 2013, \mnras, 431, 3550, \dodoi{10.1093/mnras/stt439}

\bibitem[{{G{\"o}tz} {et~al.}(2014){G{\"o}tz}, {Laurent}, {Antier}, {Covino},
  {D'Avanzo}, {D'Elia}, \& {Melandri}}]{2014MNRAS.444.2776G}
{G{\"o}tz}, D., {Laurent}, P., {Antier}, S., {et~al.} 2014, \mnras, 444, 2776,
  \dodoi{10.1093/mnras/stu1634}

\bibitem[{{Gubitosi} {et~al.}(2009){Gubitosi}, {Pagano}, {Amelino-Camelia},
  {Melchiorri}, \& {Cooray}}]{2009JCAP...08..021G}
{Gubitosi}, G., {Pagano}, L., {Amelino-Camelia}, G., {Melchiorri}, A., \&
  {Cooray}, A. 2009, \jcap, 8, 021, \dodoi{10.1088/1475-7516/2009/08/021}

\bibitem[{{Gupta} {et~al.}(2024){Gupta}, {Pandey}, {Gupta}, {Chattopadhayay},
  {Bhattacharya}, {Bhalerao}, {Castro-Tirado}, {Valeev}, {Ror}, {Sharma},
  {Racusin}, {Aryan}, {Iyyani}, \& {Vadawale}}]{2024arXiv240613755G}
{Gupta}, R., {Pandey}, S.~B., {Gupta}, S., {et~al.} 2024, \apj, 972, 166,
  \dodoi{10.3847/1538-4357/ad5a92}

\bibitem[{{He} \& {Ma}(2022)}]{2022Univ....8..323H}
{He}, P., \& {Ma}, B.-Q. 2022, Universe, 8, 323,
  \dodoi{10.3390/universe8060323}

\bibitem[{{Jacobson} {et~al.}(2004){Jacobson}, {Liberati}, {Mattingly}, \&
  {Stecker}}]{2004PhRvL..93b1101J}
{Jacobson}, T., {Liberati}, S., {Mattingly}, D., \& {Stecker}, F.~W. 2004,
  Physical Review Letters, 93, 021101, \dodoi{10.1103/PhysRevLett.93.021101}

\bibitem[{{Kislat} \& {Krawczynski}(2017)}]{2017PhRvD..95h3013K}
{Kislat}, F., \& {Krawczynski}, H. 2017, \prd, 95, 083013,
  \dodoi{10.1103/PhysRevD.95.083013}

\bibitem[{{Kosteleck{\'y}} \& {Mewes}(2001)}]{2001PhRvL..87y1304K}
{Kosteleck{\'y}}, V.~A., \& {Mewes}, M. 2001, Physical Review Letters, 87,
  251304, \dodoi{10.1103/PhysRevLett.87.251304}

\bibitem[{{Kosteleck{\'y}} \& {Mewes}(2006)}]{2006PhRvL..97n0401K}
---. 2006, Physical Review Letters, 97, 140401,
  \dodoi{10.1103/PhysRevLett.97.140401}

\bibitem[{{Kosteleck{\'y}} \& {Mewes}(2007)}]{2007PhRvL..99a1601K}
---. 2007, Physical Review Letters, 99, 011601,
  \dodoi{10.1103/PhysRevLett.99.011601}

\bibitem[{{Kosteleck{\'y}} \& {Mewes}(2013)}]{2013PhRvL.110t1601K}
---. 2013, Physical Review Letters, 110, 201601,
  \dodoi{10.1103/PhysRevLett.110.201601}

\bibitem[{{Kosteleck{\'y}} \& {Russell}(2011)}]{2011RvMP...83...11K}
{Kosteleck{\'y}}, V.~A., \& {Russell}, N. 2011, Reviews of Modern Physics, 83,
  11, \dodoi{10.1103/RevModPhys.83.11}

\bibitem[{{Kosteleck{\'y}} \& {Samuel}(1989)}]{1989PhRvD..39..683K}
{Kosteleck{\'y}}, V.~A., \& {Samuel}, S. 1989, \prd, 39, 683,
  \dodoi{10.1103/PhysRevD.39.683}

\bibitem[{{Laurent} {et~al.}(2011){Laurent}, {G{\"o}tz}, {Bin{\'e}truy},
  {Covino}, \& {Fernandez-Soto}}]{2011PhRvD..83l1301L}
{Laurent}, P., {G{\"o}tz}, D., {Bin{\'e}truy}, P., {Covino}, S., \&
  {Fernandez-Soto}, A. 2011, \prd, 83, 121301,
  \dodoi{10.1103/PhysRevD.83.121301}

\bibitem[{{Lazzati}(2006)}]{2006NJPh....8..131L}
{Lazzati}, D. 2006, New Journal of Physics, 8, 131,
  \dodoi{10.1088/1367-2630/8/8/131}

\bibitem[{{Li} {et~al.}(2009){Li}, {Mavromatos}, {Nanopoulos}, \&
  {Xie}}]{2009PhLB..679..407L}
{Li}, T., {Mavromatos}, N.~E., {Nanopoulos}, D.~V., \& {Xie}, D. 2009, Physics
  Letters B, 679, 407, \dodoi{10.1016/j.physletb.2009.07.062}

\bibitem[{{Lien} {et~al.}(2016){Lien}, {Barthelmy}, {Cenko}, {Cummings},
  {Gehrels}, {Krimm}, {Markwardt}, {Palmer}, {Sakamoto}, {Stamatikos}, \&
  {Ukwatta}}]{2016GCN.19648....1L}
{Lien}, A.~Y., {Barthelmy}, S.~D., {Cenko}, S.~B., {et~al.} 2016, GRB
  Coordinates Network, 19648, 1

\bibitem[{{Lin} {et~al.}(2016){Lin}, {Li}, \& {Chang}}]{2016MNRAS.463..375L}
{Lin}, H.-N., {Li}, X., \& {Chang}, Z. 2016, \mnras, 463, 375,
  \dodoi{10.1093/mnras/stw2007}

\bibitem[{{Magueijo} \& {Smolin}(2002)}]{2002PhRvL..88s0403M}
{Magueijo}, J., \& {Smolin}, L. 2002, \prl, 88, 190403,
  \dodoi{10.1103/PhysRevLett.88.190403}

\bibitem[{{Malesani} {et~al.}(2016){Malesani}, {de Ugarte Postigo}, {de
  Pasquale}, {Kann}, {Cano}, {Perley}, {Izzo}, {Thoene}, {Butler}, {Watson},
  {Kutyrev}, {Lee}, {Richer}, {Fox}, {Prochaska}, {Bloom}, {Cucchiara},
  {Troja}, {Littlejohns}, {Ramirez-Ruiz}, {de Diego}, {Georgiev}, {Gonzalez},
  {Roman-Zuniga}, {Gehrels}, {Moseley}, {Capone}, {Golkhou}, \&
  {Toy}}]{2016GCN.19708....1M}
{Malesani}, D., {de Ugarte Postigo}, A., {de Pasquale}, M., {et~al.} 2016, GRB
  Coordinates Network, 19708, 1

\bibitem[{{Mattingly}(2005)}]{2005LRR.....8....5M}
{Mattingly}, D. 2005, Living Reviews in Relativity, 8, 5,
  \dodoi{10.12942/lrr-2005-5}

\bibitem[{{Mitrofanov}(2003)}]{2003Natur.426Q.139M}
{Mitrofanov}, I.~G. 2003, \nat, 426, 139, \dodoi{10.1038/426139a}

\bibitem[{{Myers} \& {Pospelov}(2003)}]{2003PhRvL..90u1601M}
{Myers}, R.~C., \& {Pospelov}, M. 2003, \prl, 90, 211601,
  \dodoi{10.1103/PhysRevLett.90.211601}

\bibitem[{{Nava} {et~al.}(2012){Nava}, {Salvaterra}, {Ghirlanda}, {Ghisellini},
  {Campana}, {Covino}, {Cusumano}, {D'Avanzo}, {D'Elia}, {Fugazza}, {Melandri},
  {Sbarufatti}, {Vergani}, \& {Tagliaferri}}]{2012MNRAS.421.1256N}
{Nava}, L., {Salvaterra}, R., {Ghirlanda}, G., {et~al.} 2012, \mnras, 421,
  1256, \dodoi{10.1111/j.1365-2966.2011.20394.x}

\bibitem[{{Piran}(1999)}]{1999PhR...314..575P}
{Piran}, T. 1999, \physrep, 314, 575, \dodoi{10.1016/S0370-1573(98)00127-6}

\bibitem[{{Planck Collaboration} {et~al.}(2020)}]{2020AA...641A...6P}
{Planck Collaboration}, {et~al.} 2020, \aap, 641, A6,
  \dodoi{10.1051/0004-6361/201833910}

\bibitem[{{Rees} \& {Meszaros}(1992)}]{1992MNRAS.258P..41R}
{Rees}, M.~J., \& {Meszaros}, P. 1992, \mnras, 258, 41,
  \dodoi{10.1093/mnras/258.1.41P}

\bibitem[{{Roberts}(2016)}]{2016GCN.19224....1R}
{Roberts}, O.~J. 2016, GRB Coordinates Network, 19224, 1

\bibitem[{{Sharma} {et~al.}(2019){Sharma}, {Iyyani}, {Bhattacharya},
  {Chattopadhyay}, {Rao}, {Aarthy}, {Vadawale}, {Mithun}, {Bhalerao}, {Ryde},
  \& {Pe'er}}]{2019ApJ...882L..10S}
{Sharma}, V., {Iyyani}, S., {Bhattacharya}, D., {et~al.} 2019, \apjl, 882, L10,
  \dodoi{10.3847/2041-8213/ab3a48}

\bibitem[{{Stanbro} \& {Meegan}(2016)}]{2016GCN.19835....1S}
{Stanbro}, M., \& {Meegan}, C. 2016, GRB Coordinates Network, 19835, 1

\bibitem[{{Stecker}(2011)}]{2011APh....35...95S}
{Stecker}, F.~W. 2011, Astroparticle Physics, 35, 95,
  \dodoi{10.1016/j.astropartphys.2011.06.007}

\bibitem[{{Tasson}(2014)}]{2014RPPh...77f2901T}
{Tasson}, J.~D. 2014, Reports on Progress in Physics, 77, 062901,
  \dodoi{10.1088/0034-4885/77/6/062901}

\bibitem[{{Toma} {et~al.}(2009){Toma}, {Sakamoto}, {Zhang}, {Hill},
  {McConnell}, {Bloser}, {Yamazaki}, {Ioka}, \&
  {Nakamura}}]{2009ApJ...698.1042T}
{Toma}, K., {Sakamoto}, T., {Zhang}, B., {et~al.} 2009, \apj, 698, 1042,
  \dodoi{10.1088/0004-637X/698/2/1042}

\bibitem[{{Toma} {et~al.}(2012){Toma}, {Mukohyama}, {Yonetoku}, {Murakami},
  {Gunji}, {Mihara}, {Morihara}, {Sakashita}, {Takahashi}, {Wakashima},
  {Yonemochi}, \& {Toukairin}}]{2012PhRvL.109x1104T}
{Toma}, K., {Mukohyama}, S., {Yonetoku}, D., {et~al.} 2012, \prl, 109, 241104,
  \dodoi{10.1103/PhysRevLett.109.241104}

\bibitem[{{Vasileiou} {et~al.}(2013){Vasileiou}, {Jacholkowska}, {Piron},
  {Bolmont}, {Couturier}, {Granot}, {Stecker}, {Cohen-Tanugi}, \&
  {Longo}}]{2013PhRvD..87l2001V}
{Vasileiou}, V., {Jacholkowska}, A., {Piron}, F., {et~al.} 2013, \prd, 87,
  122001, \dodoi{10.1103/PhysRevD.87.122001}

\bibitem[{{Wei} \& {Gao}(2003)}]{2003MNRAS.345..743W}
{Wei}, D.~M., \& {Gao}, W.~H. 2003, \mnras, 345, 743,
  \dodoi{10.1046/j.1365-8711.2003.06971.x}

\bibitem[{{Wei}(2019)}]{2019MNRAS.485.2401W}
{Wei}, J.-J. 2019, \mnras, 485, 2401, \dodoi{10.1093/mnras/stz594}

\bibitem[{{Wei} \& {Wu}(2019)}]{2019PhRvD..99j3012W}
{Wei}, J.-J., \& {Wu}, X.-F. 2019, \prd, 99, 103012,
  \dodoi{10.1103/PhysRevD.99.103012}

\bibitem[{{Wei} \& {Wu}(2020)}]{2020EPJP..135..527W}
---. 2020, European Physical Journal Plus, 135, 527,
  \dodoi{10.1140/epjp/s13360-020-00554-x}

\bibitem[{{Wei} \& {Wu}(2021)}]{2021FrPhy..1644300W}
---. 2021, Frontiers of Physics, 16, 44300, \dodoi{10.1007/s11467-021-1049-x}

\bibitem[{Wei \& Wu(2022)}]{Wei2022}
Wei, J.-J., \& Wu, X.-F. 2022, Tests of Lorentz Invariance (Singapore: Springer
  Nature Singapore), 1--30, \dodoi{10.1007/978-981-16-4544-0_132-1}

\bibitem[{{Yang} {et~al.}(2017){Yang}, {Zou}, {Zhang}, {Liao}, \&
  {Lei}}]{2017MNRAS.469L..36Y}
{Yang}, C., {Zou}, Y.-C., {Zhang}, Y.-Y., {Liao}, B., \& {Lei}, W.-H. 2017,
  \mnras, 469, L36, \dodoi{10.1093/mnrasl/slx045}

\bibitem[{{Yonetoku} {et~al.}(2004){Yonetoku}, {Murakami}, {Nakamura},
  {Yamazaki}, {Inoue}, \& {Ioka}}]{2004ApJ...609..935Y}
{Yonetoku}, D., {Murakami}, T., {Nakamura}, T., {et~al.} 2004, \apj, 609, 935,
  \dodoi{10.1086/421285}

\bibitem[{{Yonetoku} {et~al.}(2012){Yonetoku}, {Murakami}, {Gunji}, {Mihara},
  {Toma}, {Morihara}, {Takahashi}, {Wakashima}, {Yonemochi}, {Sakashita},
  {Toukairin}, {Fujimoto}, \& {Kodama}}]{2012ApJ...758L...1Y}
{Yonetoku}, D., {Murakami}, T., {Gunji}, S., {et~al.} 2012, \apjl, 758, L1,
  \dodoi{10.1088/2041-8205/758/1/L1}

\bibitem[{{Zhou} {et~al.}(2021){Zhou}, {Yi}, {Wei}, \&
  {Wu}}]{2021Galax...9...44Z}
{Zhou}, Q.-Q., {Yi}, S.-X., {Wei}, J.-J., \& {Wu}, X.-F. 2021, Galaxies, 9, 44,
  \dodoi{10.3390/galaxies9020044}

\end{thebibliography}

\end{document}